\documentclass[twocolumn,pre,nobibnotes,altaffilletter,amsmath,amssymb,amsfonts]{revtex4-1}
\usepackage[latin1]{inputenc}
\usepackage{graphicx}
\usepackage{amsmath}
\usepackage{latexsym}
\usepackage{epsfig}
\usepackage{subfig}
\usepackage{subfloat}
\usepackage{color}
\makeindex

\begin{document}

\title{Observable-dependence of the effective temperature in off-equilibrium diatomic molecular liquids.}

\author{Andrea Saverio Ninarello$^1$}
\author{Nicoletta Gnan$^2$}\email{nicoletta.gnan@roma1.infn.it} 
\author{Francesco Sciortino$^1$}
\affiliation{$^1$ Dipartimento di Fisica, "Sapienza" Università di Roma, Piazzale A. Moro 2, 00186 Roma}
\affiliation{$^2$ CNR-ISC Uos "Sapienza", Piazzale A. Moro 2, 00186 Roma}

\date{today}

\begin{abstract}
  
We discuss the observable-dependence of the effective temperature $T_{eff}$, defined via the fluctuation-dissipation relation, of an out-of-equilibrium system composed by homonuclear dumbbell molecules. $T_{eff}$ is calculated by evaluating the fluctuation and the response  for two observables associated respectively to translational and  to rotational degrees of freedom,  following a sudden temperature quench. We repeat our calculation for different dumbbell elongations $\zeta$. At high elongations ($\zeta > 0.4 $),
we find the same  $T_{eff}$  for the two  observables.
At low elongations ($\zeta \leq 0.4$), only for very deep quenches $T_{eff}$ coincides. The  observable-dependence of $T_{eff}$  for low elongations and shallow quenches stresses the importance 
 of a strong coupling between orientational and translational variables 
for a consistent definition of  the effective temperature in  glassy systems.
  
\end{abstract}


\maketitle
\section*{Introduction}

Understanding  the off-equilibrium state of matter,   central in   glass science~\cite{GOETZE2009,BINDER2005,WOLYNES2012,Leuzzi2009,Berthier2011,Biroli2013,Kurchan2005,Ediger1996}, is a challenging task.  The difficulty arises from  the lack of a thermodynamic approach capable of describing glasses.
Indeed, in addition to temperature and pressure, other macroscopic observables are requested to uniquely describe the
state of the system. 
When a liquid is brought off-equilibrium, e.g. by quickly lowering the bath temperature $T_{bath}$, dynamic properties  depend on the observation time $t'$, i.e. the system ages.   Correlation functions retain their two step
decay behavior characteristic of supercooled equilibrium states, but show a  $t'$ dependence mainly  of the $\alpha$-relaxation  process:  the structural relaxation time, controlled by the slow modes, increases with the observation time. Aging thus affects  the long-time dynamics of particles. The separation of time scales reflects  particles vibration around their average position at short times and the diffusional process at long times. As thus,  the fast vibrational dynamics  equilibrates to $T_{bath}$ and does not significantly changes with aging. 

It has been suggested that  the evolution of the correlation function with $t'$ could be interpreted as
 arising from the slow  thermalization  of the structural  degrees of freedom associated to the  $\alpha$-process.
 In this respect, one could add to the state variable one or more additional temperature of the slow modes.
 While it is clear from   experiments based on the Kovacs protocol~ \cite{Kovacs1963, Angell2000}, reproduced also in silico~\cite{Mossa2004,Saika2004},   that under strong quenching, more than one  additional temperature is needed,
 other protocols  are compatible with the idea that one single   additional parameter is sufficient to characterize the aging system. In this picture, aging corresponds to the evolution of this additional temperature  toward $T_{bath}$.
In the theoretical framework developed by Cugliandolo and Kurchan \cite{Cugliandolo1997, Cugliandolo2000}, the additional temperature reveals its presence in the violation of 
the  fluctuation-dissipation theorem (FDT). 
In equilibrium the response  of a system to an
external weak perturbation is  linearly related to a suitable correlation function (and the coefficient of linearity is $T_{bath}$).  In aging, the fluctuation-dissipation theorem is generalized to account for 
the presence of an additional temperature, introducing a  fluctuation-dissipation (FD) ratio.  Given the correlation function $C_{AB}(t,t')=<A(t)B(t')>_0$ of two  variables $A(t)$ and $B(t)$ in an unperturbed system, and the integrated response $\chi_{A,B}(t, t')= \int_{t'}^t  \frac{\delta<A(t)>_h}{\delta h(s)} ds \rvert_{h\rightarrow 0}$ to a perturbation field $h$ applied at a time $t>t'$ (and coupled to the observable $B(t)$), the two functions are related by the expression~\cite{Leuzzi2009,Crisanti2003}

\begin{equation}
\frac{ \partial C_{A,B} (t,t')}{\partial t'} = 
-\frac{T_{bath}}{X_{A,B}(t,t')} \frac{\partial \chi_{A,B}(t,t')}{\partial t'}
\label{partialFDR}
\end{equation}
where 
\begin{equation}
X_{A,B}(t,t') = -\frac{1}{T_{bath}} \frac{\partial \chi_{A,B}(t,t')}{\partial C_{A,B}(t,t')}
\bigg|_{\substack{ t'=fixed \\ C_{A,B}(t,t')=const}}
\end{equation}
is the fluctuation-dissipation (FD) ratio. The  equilibrium results are recovered when  $X_{A,B}(t,t')=1$ and the FDT is satisfied.

When the system is out of equilibrium the fluctuation and the integrated response are differently
related  at short and long time-scales. For short time-scales

\begin{equation}
|X_{A,B}(t,t')| = 1 \ \ \ \ \ \ \ \ \text{if} \ \ (t-t')/t' \ll 1,
\end{equation}
i.e. the aging system behaves as if it were in equilibrium,  while at long times

\begin{equation}
|X_{A,B}(t,t')| = \frac{T_{bath}}{T_{eff}(t,t')} \ \ \ \ \ \ \ \ \text{if}
\ \ (t-t')/t' \gg 1,
\end{equation}

\noindent where $T_{eff}$, named \emph{effective temperature}, has been interpreted as the
temperature of the slow degrees of freedom.  Thus, while the vibrational dynamics is instantaneously in equilibrium at $T_{bath}$ after a temperature quench, the slow modes  can be thought to be in \emph{quasi} equilibrium \cite{Crisanti2003,Cugliandolo2011} at $T_{eff}(t') > T_{bath}$.  The $t'$ dependence clarify that $T_{eff}$ progressively evolves in time toward $T_{bath}$. Hence the FDR in structural glasses leads naturally to a rigorous definition of $T_{eff}$ that can be  evaluated  estimating the FD ratio. 
Evidence of a two-slope FDT in aging systems have been largely provided by numerical simulations of supercooled liquids following temperature or pressure changes~\cite{Parisi1997,Barrat1999a, Kob2000,Leonardo2000,Sciortino2001,Berthier2007MC,Gnan2010,Gnan2013}.
For these cases, it has been suggested that $T_{eff}$ coincides with the 
temperature that a thermometer, weakly coupled to the system, would measure in an aging glass if its internal time-scale is equal to the time-scale of the slow processes in the glass~\cite{Cugliandolo1997,Kurchan2005}. In addition, in some simple molecular glass models, $T_{eff}$ has been shown to coincide with an {\it internal} temperature independently obtained from an extended thermodynamic framework based on the potential energy landscape (PEL) approach.~\cite{Sciortino2001,Sciortino2005,Gnan2013}.
Finally, while the assumption of one single additional parameter has been questioned in the past~\cite{Kovacs1963, Angell2000,Mossa2004}, the hypothesis of one single  $T_{eff}$  appears to be particularly  appropriate  in some class of systems~\cite{Pedersen2008,Bailey2008,Gnan2010}. 
For these systems  the effective temperature could enter, along with the other equilibrium thermodynamic parameters, in a two-temperatures thermodynamic approach in which a separation of time scales is assumed~\cite{Leuzzi2009}. 

%RELATION WITH ENTROPY PROD AND FT [ZAMPONI,SELLITTO]?
Although the $T_{eff}$ defined from the FDR has been proven to be a valuable concept in several cases, few studies have focused on its
observable dependence. Indeed if $T_{eff}$ can be interpreted as a genuine thermodynamic parameter it should coincide when measured for different observables on the same time-scale. There is no experimental evidence of the observable independence of $T_{eff}$ in glassy systems and only a few  theoretical~ \cite{Cugliandolo2000} or  numerical investigations~\cite{Berthier2004Shear,Berthier2007a}. 
 In the numerical investigation of Lennard-Jones (LJ) atomic liquids~\cite{Berthier2004Shear,Berthier2007a},  it was found that the FD ratio built from calculating the fluctuations and the response of the density at different wave vectors results in the same $T_{eff}$. However such result is limited to a single system and all the observables employed in the study are related to translational degrees of freedom.  At odd, the scarce experimental works have focused on 
  molecular glasses \cite{Grigera1999,Buisson2005,Reiser2011,Oukris2010} and related to the observation of fluctuations and responses of molecules orientation.

In this study we scrutinize the observable-dependence of $T_{eff}$, by measuring intrinsically different (i.e. translational and rotational)  observables. We study the out-of-equilibrium behaviour of a system of homonuclear molecules with
different elongations, probing both translational and rotational degrees of freedom. We show that various
scenarios can occur as a function of the molecules elongation and that distinct observables can be found in distinct
aging conditions during the same temperature quench. This   results in a partial decoupling of the translational and rotational degrees of freedom and hence to dissimilar FD ratios.

\section*{Models and methods}
We perform Metropolis Monte Carlo (MC) simulations of a binary mixture $80:20$ of $N=500$ homonuclear dumbbell molecules interacting via a cut-and-shifted LJ potential~\cite{Chong2005},

\begin{equation}
V_{\alpha\beta}(r)=4\varepsilon_{\alpha\beta}\left[\left( \frac{\sigma_{\alpha\beta}}{r}\right)^{12}-\left( \frac{\sigma_{\alpha\beta}}{r}\right)^{6} +A_{0}+A_{1}\frac{r}{\sigma_{\alpha\beta}}\right].
\end{equation}
\noindent Here $r$ is the center-to-center distance of two atoms belonging to molecules of type $\alpha$ and $\beta$ (with $\alpha,\beta \in{A,B}$).  $\sigma_{\alpha\beta}$ and 
$\varepsilon_{\alpha\beta}$ quantify the characteristic size and  depth of the interaction potential. The two constants $A_0$ and $A_1$ are set to guarantee the continuity
of the potential and of its derivative at the cut-off distance $c=2.5$  (in units of $\sigma_{\alpha\beta})$.
 From such two constraints one finds that $A_0=c^{-6}(7-13c^{-7})$ and $A_1=6c^{-7}(2c^{-6}-1)$.

\begin{figure}
\begin{center}
\includegraphics[width = 0.4\textwidth]{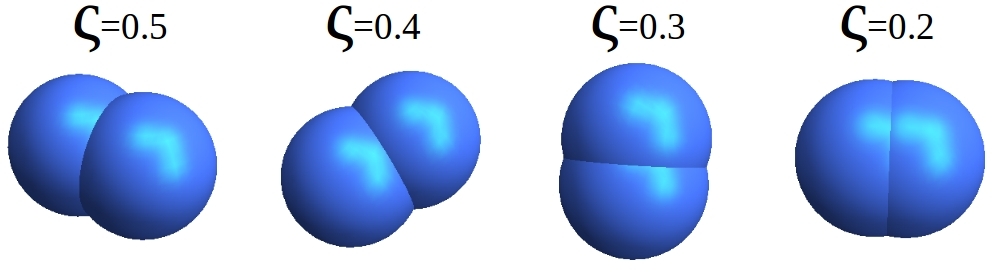}
\caption{Snapshots of the dumbbell molecules as a function of the elongation}.
\label{Dumb}
\end{center}
\end{figure}

Following Ref.~\cite{Chong2005}
we set the interaction parameters to $\sigma_{AA}=1.0$, $\sigma_{AB}=0.8$, $\sigma_{BB}=0.88$, $\varepsilon_{AA}=1.0$, $\varepsilon_{AB}=1.5$ and $\varepsilon_{BB}=0.5$, which correspond to the values first introduced by Kob and Andersen \cite{kobAndersen1995,kobAndersen19952} for a binary mixture of Lennard-Jones atoms with the aim to prevent the mixture from crystallizing at high densities. In the following, all the parameters will be expressed in reduced units, selecting $\varepsilon_{AA}$ and $\sigma_{AA}$ as units of energy and distance. We also set $k_B=1$.
 In the the case of dumbbell molecules, the total packing fraction $\phi_{tot} = \phi_{AA} + \phi_{BB}$ can be defined as \cite{Chong2005}:

\begin{equation}\phi_{\alpha \alpha} = \frac{\pi}{6} \rho_{\alpha \alpha} \sigma^3_{\alpha \alpha} 
\left( 1+ \frac{3}{2}\zeta - \frac{1}{2} \zeta^3
\right) ,\ \  0\leq \zeta \leq 1 ,\ \ \alpha\in\{A,B\}
\end{equation}
\noindent where $\rho_{\alpha \alpha} = \frac{N_{\alpha}}{V}$ is the number density (namely the number of $\alpha$  dumbbells  over the volume), while $\zeta$ is the elongation, i.e. the bond length $l_{\alpha\alpha}$ between the centers of the two atoms forming the molecule, expressed in units of
$\sigma_{\alpha\alpha}$ (i.e. $\zeta=l_{AA}/\sigma_{AA}=l_{BB}/\sigma_{BB}$). 
A series of snapshots showing how much the shape of the molecules
changes when decreasing $\zeta$ are reported in Fig. \ref{Dumb}.
In our study 
we fix the packing fraction to $\phi=0.708$ in order to compare our equilibrium results with previous Molecular Dynamics (MD) studies at different elongations \cite{Chong2005,Moreno2005}.
In the following, we define a MC step as $N$ attempts
to translate and rotate  of a random quantity a randomly selected dumbbell. 

Since our work concerns the study of dynamical quantities obtained from a stochastic dynamics (MC), it is legitimate to ask
whether the Metropolis algorithm reproduces a long-time dynamical behavior  similar to the one
observed with   Newtonian dynamics~\cite{Chong2005,Moreno2005} for the same system.
This question has been previously addressed for supercooled 
 binary mixtures of LJ atoms~\cite{Berthier2007MC} and for models of colloidal particles~\cite{Sanz2010, Romano2011}. These studies have provided evidence that the Metropolis algorithm can gives rise to a physically relevant slow dynamics.  More precisely, it has been shown that 
 the  long-time-decay of   the self-intermediate scattering function is identical (with a proper time
 rescaling) for Newtonian, Brownian~\cite{Gleim1998} and Monte-Carlo simulations~\cite{Berthier2007MC}.  
Following a procedure similar to that presented in Ref.~\cite{Berthier2007MC}, we have verified that the long-time relaxation behavior for density and angular correlators in equilibrium LJ dumbbells is equivalent to that evaluated with Netwonian dynamics~\cite{Chong2005,Moreno2005}. A detailed discussion for the case of molecules with $\zeta=0.5$ can be found in Appendix \ref{appendix:MCvsMD}.

In order to study the FDR  we exploit a zero-field
Monte-Carlo (MC) algorithm developed and tested previously on atomic structural glasses. This algorithm  provides an unbiased measurement of the response function \cite{Berthier2007a}. The method allows us, within the same simulation,
the simultaneous measure of both the response and the correlation at different waiting times $t'$ until a fixed time $t$ from the quench, thus reducing the computational load.
Following standard works on FDT violation in glasses \cite{Crisanti2003}, we measure $T_{eff}$ by building the fluctuation-dissipation (FD) plot; it consists in  reporting $T_{bath}\chi(t,t')$ versus $C(t,t')$. If the system is characterized by a clear separation of time-scales, as in our case, the resulting parametric curve will display two slopes; the FD ratio $X=T_{bath}/T_{eff}$ can be immediately visualized being the angular coefficient of the straight line different from $X=1$. As a rule, we chose to evaluate $X$ via a linear fit of the points in the FD-plot satisfying the relation $(t-t')/t'>2$.  

In this study, we  calculate $T_{eff}$ for observables associated to translational and rotational degrees of freedom.
For  translations, we choose  $A_{TRANS}(t)=N^{-1}\sum_i\epsilon_i e^{-i \boldsymbol{k} \cdot \boldsymbol{r}_i(t)}$ and $B_{TRANS}(t)=2\sum_j \epsilon_j\cos(\textbf{k}\cdot \boldsymbol{r}_j(t))$ where the coordinates $\boldsymbol{r}_i(t)$ are the positions of the centres of mass of the molecules at time $t$ and $\epsilon_j=\pm 1$ is a bimodal variable
with zero mean which suppresses the cross terms in the correlation and the response functions~\cite{Berthier2007a}.
Thereby the  correlation function $F_s(\boldsymbol{k},t,t') \equiv \langle A_{TRANS}(t)B_{TRANS}(t')\rangle_0$ corresponds to the self-intermediate scattering function, where the zero subscript indicates the average over unperturbed trajectories.  
For  rotations we choose $A^{(l)}_{ROT}(t)=\sqrt{{\cal N}_l}N^{-1}\sum_i \epsilon_i P_l[\cos(\theta_i(t))]$
and $B^{(l)}_{ROT}(t)=\sqrt{{\cal N}_l}\sum_j\epsilon_jP_l[\cos(\theta_j(t))]$, where $\epsilon_i$ is a bimodal variable as above, $P_l$ is the Legendre polynomial of order $l$ and $\theta_i(t)=\hat{e}_i(t) \cdot \hat{x}$ is the projection of the molecular axis on the $x$-axis. In addition ${\cal N}_l$ is a normalization constant which depends on the order $l$ of the Legendre polynomial and ensures the angular correlation function $C_{l}(t,t')=\langle A^{(l)}_{ROT}(t) B^{(l)}_{ROT}(t')\rangle_{0}$ to be $1$ when $t=t'$. Specifically, $\cal{N}$$_1=\frac{1}{3}$ and $\cal{N}$$_2=\frac{1}{5}$. 
In the following we will refer to the FD-plots of the two sets of observables (translation and rotation) as the self-density and the $l$-orientation FD-plots.

\section*{Results and Discussions}

\begin{figure*}[t!]
\centering
\subfloat[]{\includegraphics[width=0.378\textwidth]{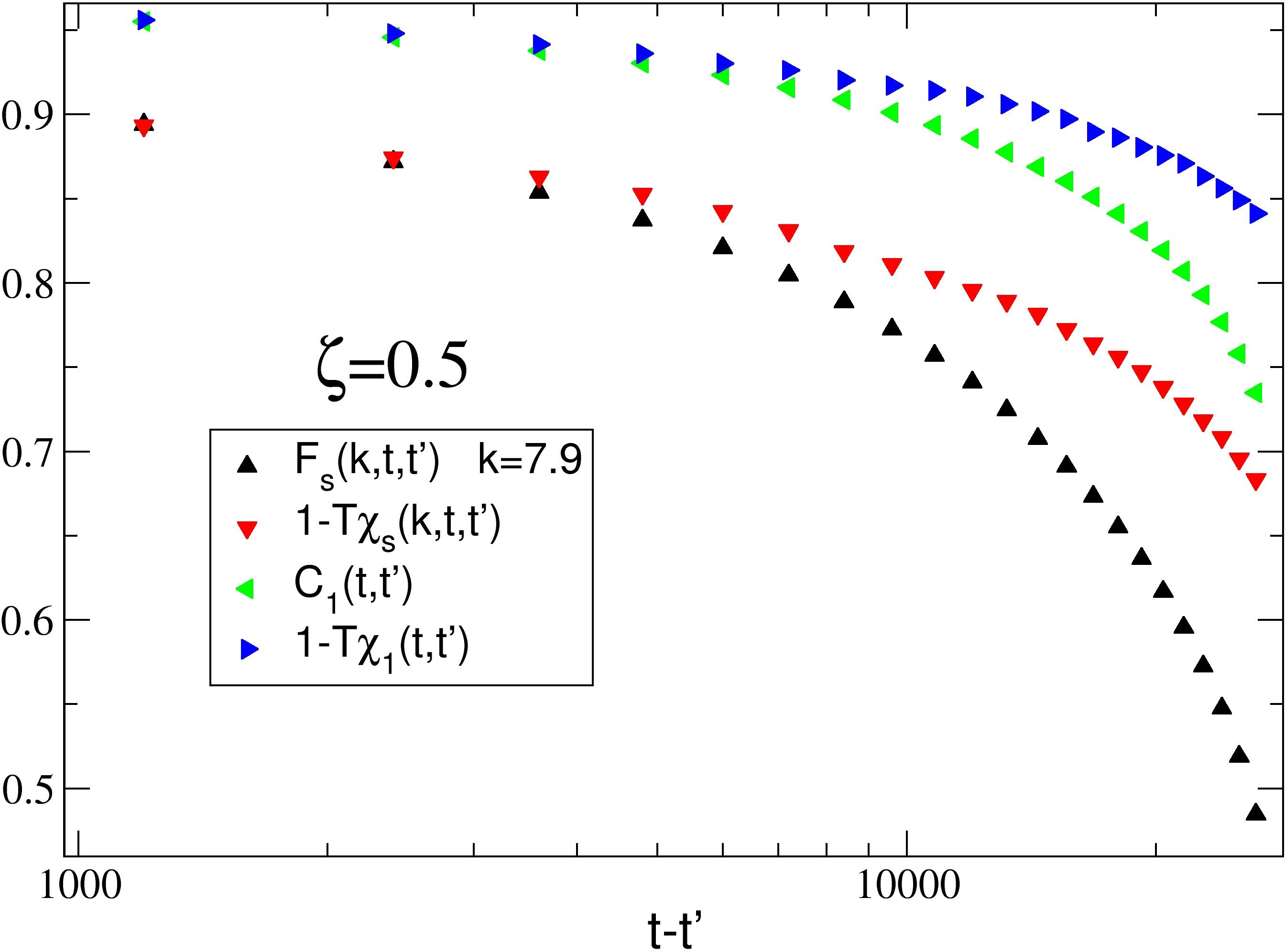}}\hspace{0.1cm}
\subfloat[]{\includegraphics[width=0.4\textwidth]{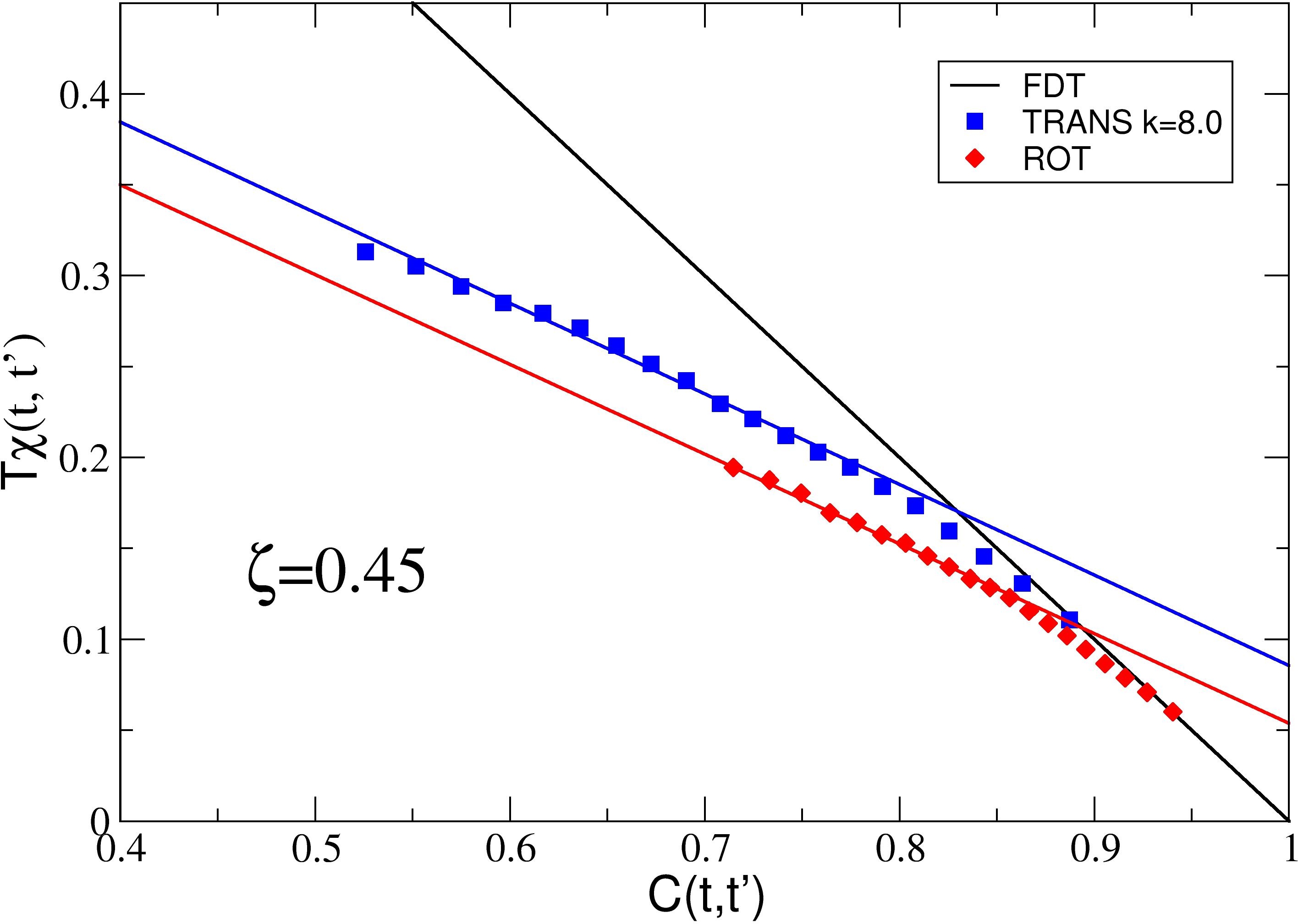}}

\subfloat[]{\includegraphics[width=0.4\textwidth]{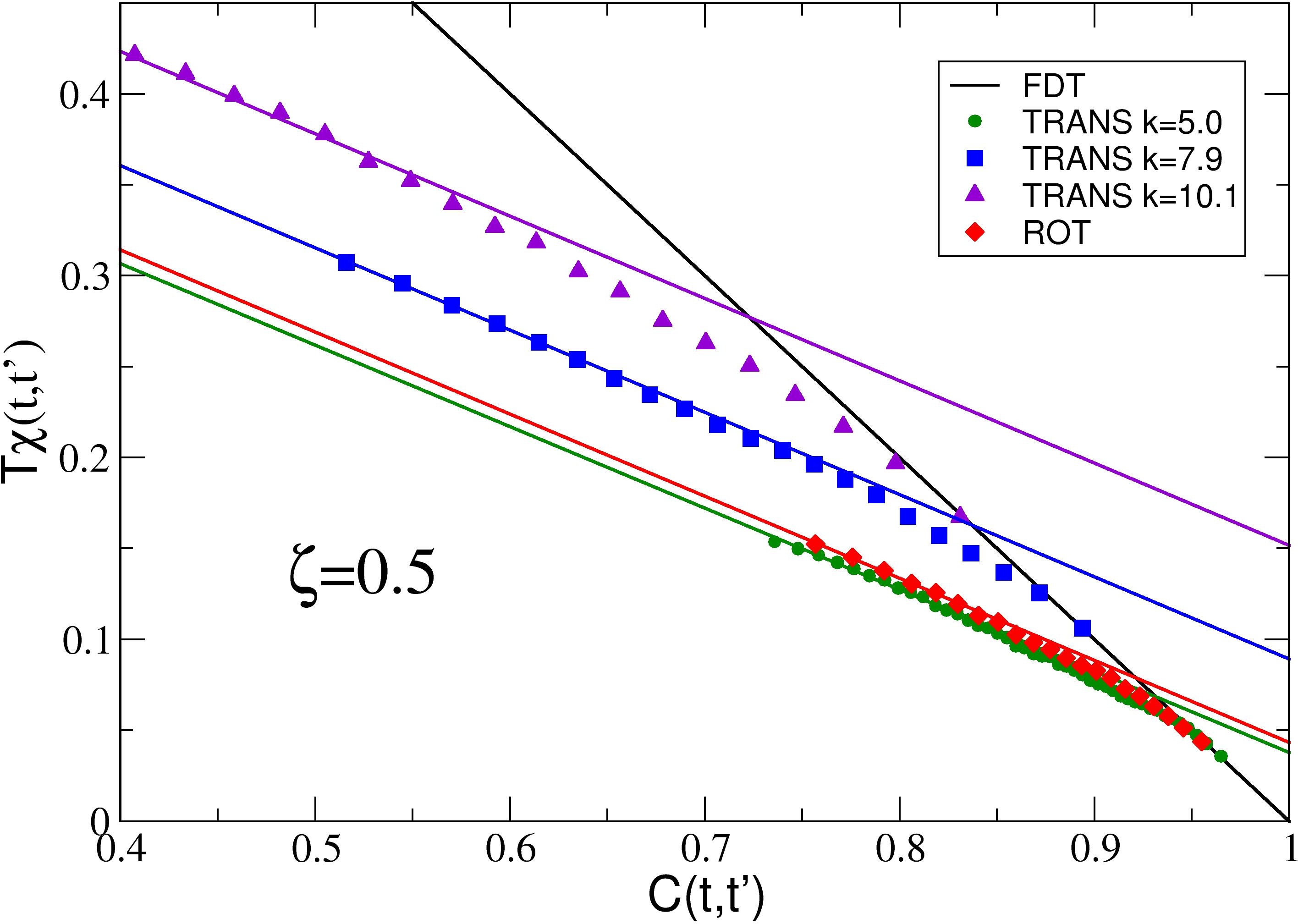}}\hspace{0.1cm}
\subfloat[]{\includegraphics[width=0.4\textwidth]{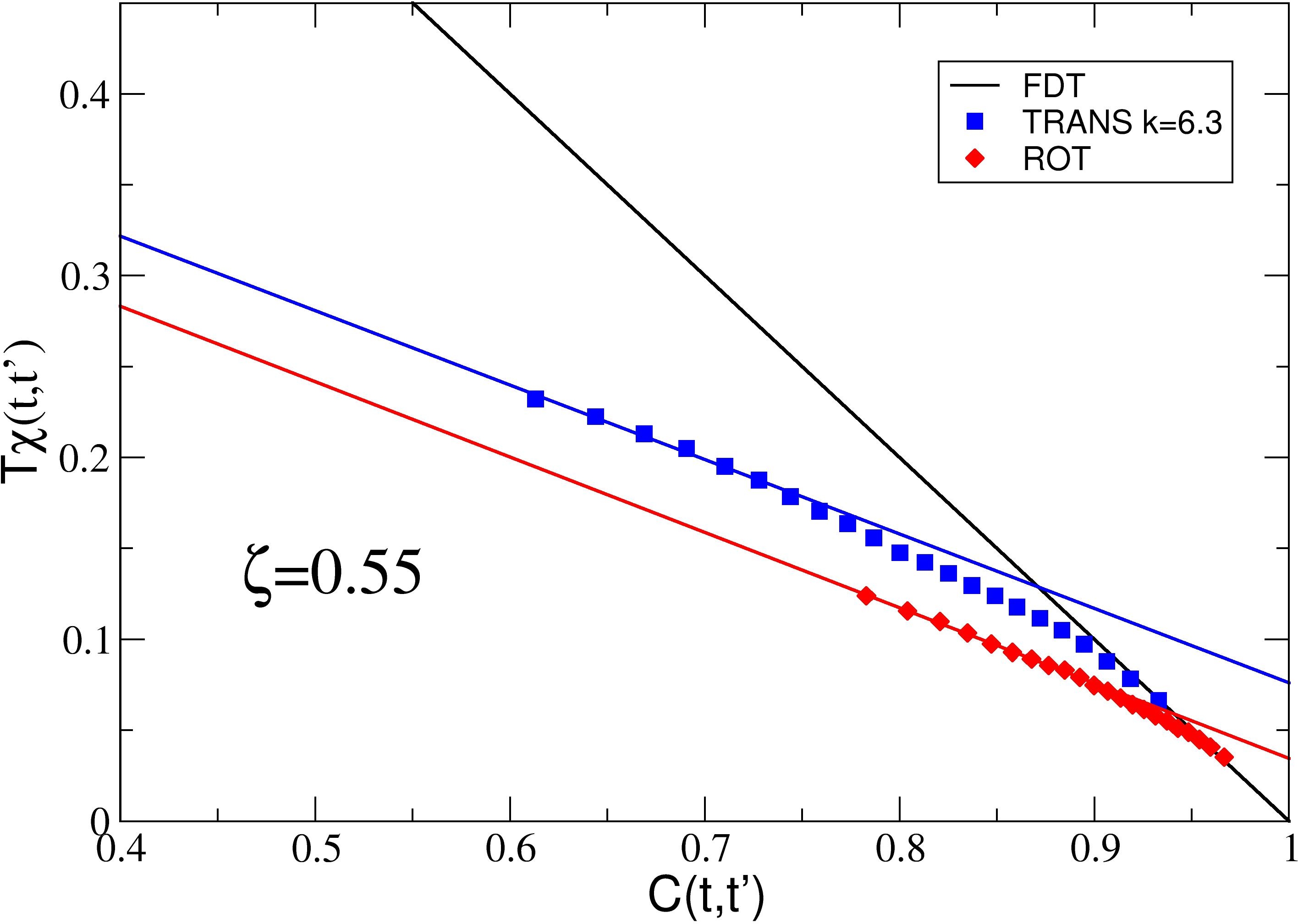}}
\caption{(a) Correlation and integrated responses as a function of $t-t'$ for the system of LJ dumbbells with elongation $\zeta=0.5$. Functions are, respectively,  $F_s(\textbf{k},t,t')$ (black up triangles), $1-T_{bath}\chi_s(\textbf{k},t,t')$(red down triangles),  $C_1(t,t')$ (green left triangles) and $1-T_{bath}\chi_1(t,t')$ (blue right triangles). Data are taken for the quench $T=6.0\rightarrow T=0.4$.
(b) Self-density (blue squares) and $l=1$-orientation (red diamonds) FD-plots for $\zeta=0.45$.
(c) Self-density FD-plots for wave vectors $|\textbf{k}|=10.1$ (violet up triangles), $|\textbf{k}|=7.9$ (blue squares) and $|\textbf{k}|=5.0$ (green circles), together with the $l=1$-orientation FD-plot (red diamonds) for LJ dumbbells with $\zeta=0.5$.
(d) Self-density (blue squares) and $l=1$-orientation (red diamonds) FD-plots for $\zeta=0.55$.
}
\label{totH}
\end{figure*}

We perform several temperature quenches by instantaneously lowering $T$ from $T=6$ to $T=0.4$ for different 
elongations and we observe the off-equilibrium evolution
of the correlators ($F_s(\textbf{k},t,t')$ and $C_1(t,t')$) and of the corresponding integrated responses
($\chi_s(\textbf{k},t,t')$ and  $\chi_1(t,t')$) up to a fixed time $t$ from the quench. This allows us to build the FD-plot
from which it is possible to extract $T_{eff}$. When not specified, we select  $|\textbf{k}| \simeq 8.0$,  that roughly corresponds to the
first peak of the static structure factor $S(k)$ of the species $A$.
We calculate $T_{eff}$ both in the high elongation region (i.e. $\zeta > 0.4$) and
in the low elongation region ($\zeta \leq 0.4)$. 
\newline

\subsubsection{High elongation region ($\zeta > 0.4$)}
We investigate the off-equilibrium dynamics of molecules with elongation $\zeta=0.45,0.5,0.55$. For each elongation we perform  trial quenches and monitor the evolution of the potential energy during aging.  We select the
smallest investigated waiting time $t'$ when  the
 potential energy   starts to show a logarithmic behavior, indicative of a  "quasi stationary" aging regime.  We find such time to be $t'=1800$ MC steps for the three elongations. Successive waiting times are separated one from the other by $1200$ MC steps. The total observation time $t$ is set to $t=29400$ MC steps.
 
We have not been able to investigate the cases  with  $\zeta>0.55$. Indeed, for these values 
 rotations are severely frozen; this implies that, when building the $l=1$-orientation FD-plot, the second slope is not well developed within our observation-time window and we cannot evaluate $T_{eff}$. A  measurement of $T_{eff}$  would require prohibitively longer times, inaccessible with our computational facilities.

For $\zeta=0.45,0.5,0.55$, we perform $20000$ independent quenches to ensemble average the correlators and the response functions requested for drawing the FD-plots.
Fig.~\ref{totH}(a)  clearly shows that, differently from equilibrium, the integrated response  (plotted as $1-T_{bath}\chi$) evolves in time more slowly than the correlation, both for 
translational and rotational observables. Figs.~\ref{totH} (b),(c),(d) show  the  three FDT violations. 
The two-time-scale separation between fast and slow modes in the model system gives rise to FD-plots characterized by two slopes.
For both translational and rotational observables,  the resulting parametric plots display the same FD ratio $X(t,t')$ and hence the same $T_{eff}$.

Fig.\ref{totH}(c) ($\zeta=0.50$)  shows also the FD-plot for the self-density  evaluated at two other different $k$ vectors. In all cases, the same slope is found, confirming that $T_{eff}$ does not depend on $k$,
in agreement with results for atomic systems\cite{Barrat1999b,Berthier2007a}. 
%The self-density parametric plot does not show significant dependence on  $\zeta$ when $k$ is fixed  (fig. \ref{totH} (b),(d)), suggesting that density responses and correlations occur at similar time and space scales at different $\zeta$. 

%In the FD-plot the intersection between the $X=1$ and $X=T_{eff}/T_{bath}$ lines  locates  the plateau height of the two-time correlation function. Noticeably, this point changes for  the rotational degrees of freedom FD-plot  on varying $\zeta$, suggesting   that the rotational decorrelation and response processes have a strong dependence on the elongation.

\subsubsection*{Low elongation region ($\zeta \leq 0.4$)}
\begin{figure*}
\centering
\subfloat[]{\includegraphics[width=0.4\textwidth]{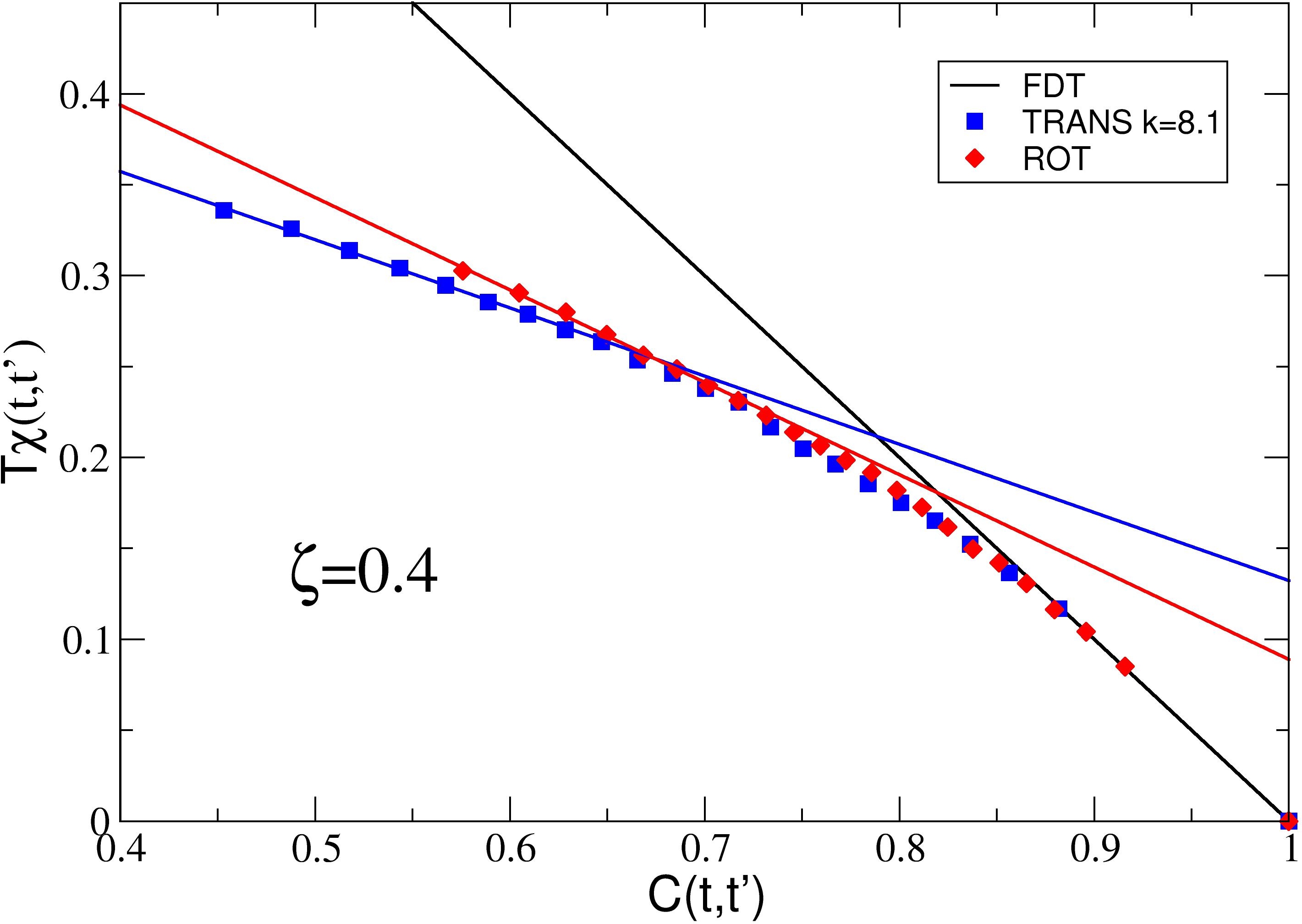}}\hspace{0.1cm}
\subfloat[]{\includegraphics[width = 0.4\textwidth]{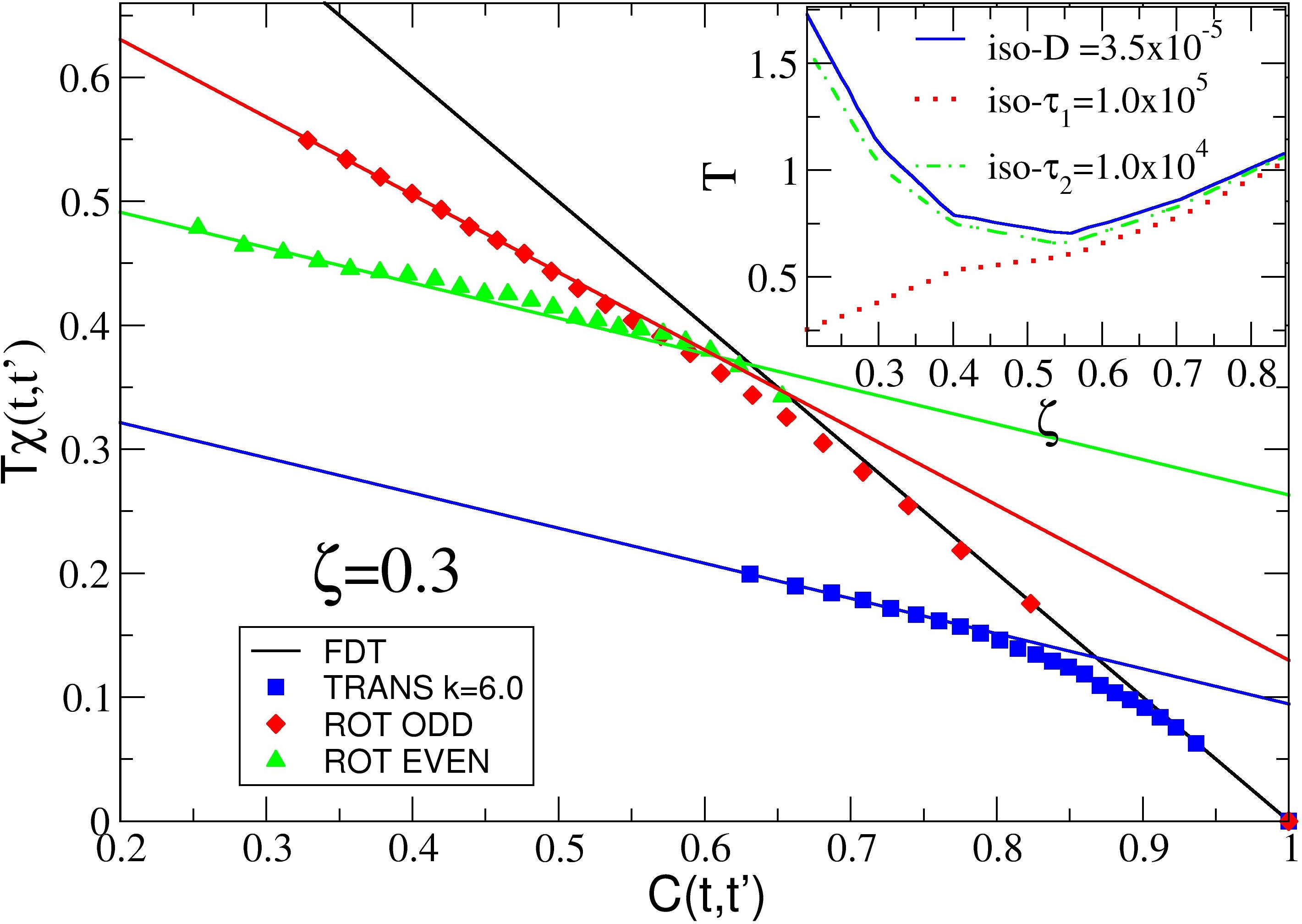}}
\centering
\caption{self-density (blue squares) and $l=1$-orientation (red diamonds) FD-plots for  dumbbells with elongation (a)$\zeta=0.4$ and (b)$\zeta=0.3$, for the quench $T=6.0\rightarrow T=0.4$. Green squares in (b) is the $l=2$-orientation FD-plot, associated to even rotations of molecules. Inset: Isodiffusivity line (thin continous line) and isochrones for the decay time $\tau_1$ of $C_1(t)$ (thin dotted line) and $\tau_2$ of $C_2(t)$ (dashed dotted line) in the $T-\zeta$ plane. Lines connect data points obtained form MD simulations. Data taken form \cite{Moreno2005}. }
\label{totL}
\end{figure*}

In this region we investigate molecules with elongations $\zeta= 0.2,0.3,0.4$. 
As for high elongations, we follow the decay of the potential energy during a quench to identify a quasi equilibrium regime. For the three elongations we set the first waiting time to $t'=1800$ MC steps, the same value used for large $\zeta$. As before, successive waiting times are separated by $1200$ MC steps and the total observation time is $t=29400$ MC steps.
We perform a number of quenches varying between $10000$ and $20000$ for each elongation and we  build the self-density and the $l=1$-orientation FD-plots, finding again the two-slope scenario. 
However if we extract $T_{eff}$ from the parametric plots, we find  two different effective temperatures for the two FD-plots. The difference between the two $T_{eff}$s  increases when decreasing $\zeta$. This is shown in Fig.\ref{totL} (a) and (b), respectively for  $\zeta=0.4$ and $\zeta=0.3$. Note that the slope of the FD-plot relative to $l=1$-orientation  is always higher than the slope relative to  the density; it follows that the $T_{eff}$  for $l=1$-orientation  is always closer to $T_{bath}$ than the  self-density $T_{eff}$. 

A hint to rationalize the previous results comes from the equilibrium dynamic behaviour of the dumbbell molecules in the supercooled regime at different $\zeta$.
Previous theoretical and numerical studies \cite{Chong2002,Moreno2005,Chong2005} have investigated the dynamic phase diagram of the dumbbells in the $T-\zeta$ plane, showing a strong dependence of the relaxation dynamics on the elongation of the molecules. For instance, numerical simulations \cite{Moreno2005} have shown that iso-diffusivity lines  are non-monotonic functions of the elongation. The behaviour of the rotations is also intriguing: depending on the degree $l$ of the Legendre polynomial employed to build $C_{l}(t)$, the observed dynamics and the isochronal iso-$\tau_l$ curves (where $\tau_l$ is the relaxation time of $C_l(t)$) can be different (inset of Fig.\ref{totL}(b)).
Specifically, for even-$l$ correlators, e.g. $C_{l=2}(t)$, the iso-$\tau_2$ curve closely follows the iso-diffusivity line in the $T-\zeta$ plane. Contrary, the iso-$\tau_1$ line of $C_1(t)$ is coupled to the iso-$D$ and to the iso-$\tau_2$ curves only at high $\zeta$, while the coupling is lost for lower $\zeta$.  In this decoupled region, 
 $C_1(t)$ relaxes  to zero significantly faster than $C_2(t)$. This peculiar behaviour has been explained in terms of hopping processes occurring in the low-elongation region. Indeed it has been observed that low-elongation molecules perform rotational flips of $180^\circ$ which change sign to $P_1$  (and to all  odd Legendre polynomial), providing the dominant contribution to the decorrelation of  $C_1(t)$.  According to Mode-Coupling Theory for monodisperse dumbbells \cite{Chong2002} a critical value for the molecules elongation $\zeta_c = 0.345$ marks the crossover between the strong-to-weak hindrance scenario described above. Simulations also suggest a value close to $\zeta_c$ for the present binary mixture of dumbbells.  These equilibrium results provide two important pieces of evidence:
 (i) that there is a cross-over between small and large $\zeta$ values, associated to a decoupling of rotation and translation for odd $l$. (ii) that the decay of the odd $l$ correlation functions for small $\zeta$ proceed much faster than the decay of the other correlators. 

Based on these findings we investigate the off-equilibrium behaviour of the $l=2$ orientation $C_{2}(t,t')$ and its integrated response $\chi_2(t,t')$ when $\zeta=0.3$. Figure \ref{totL} (b) shows the resulting $l=2$-orientation FD-plot: we find that, as for the equilibrium case, the aging behaviour of observables associated to translational and $l=2$-rotational degrees of freedom are strongly coupled.   Differently from what found for $l=1$,
the two FD-plots display now the same $T_{eff}$.

\begin{figure}
\includegraphics[width = 0.4\textwidth]{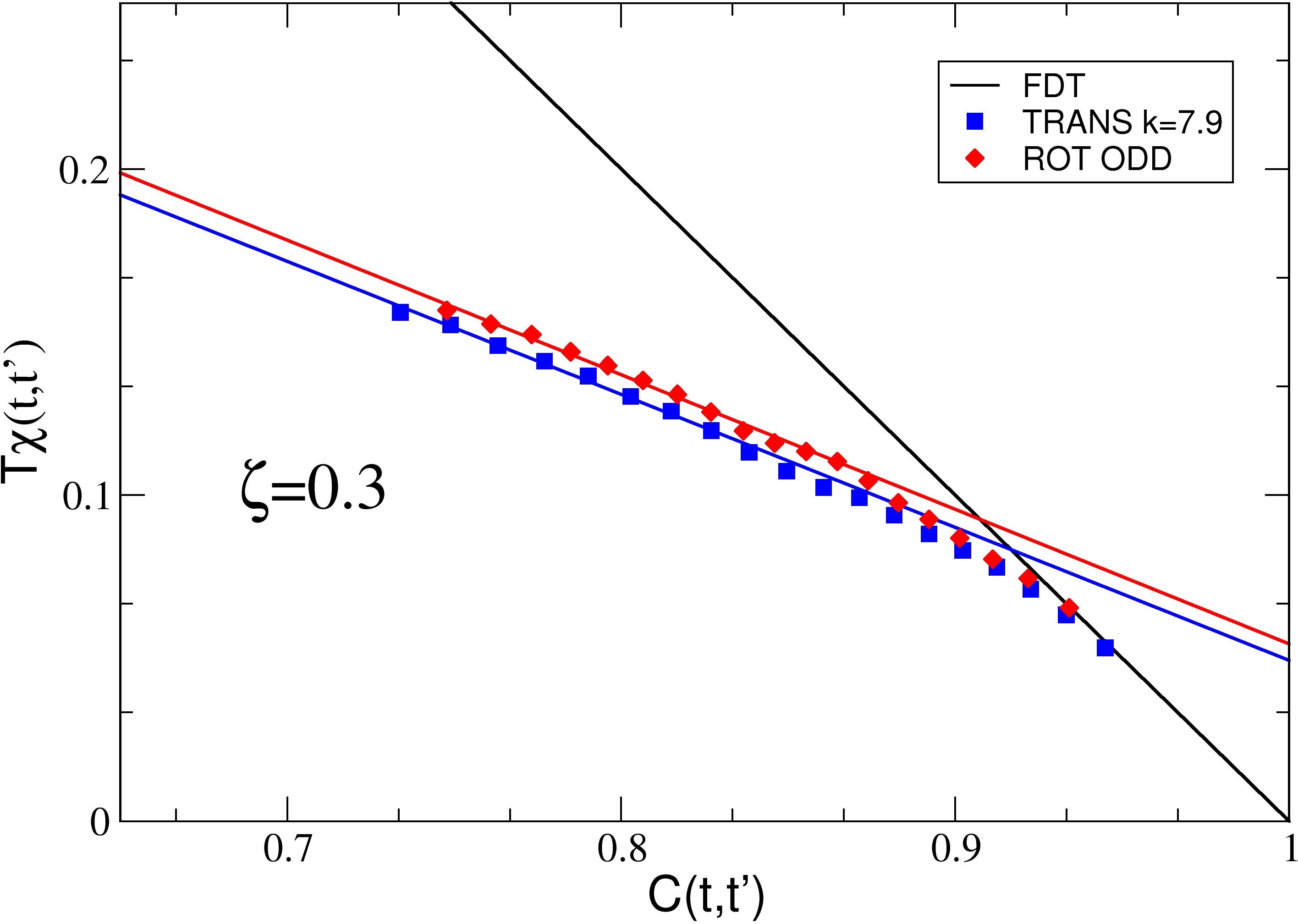}
\caption{self-density and $l=1$-orientation FD-plots for the system of dumbbells with $\zeta=0.3$ for the quench $T=6.0\rightarrow T=0.2$ }
\label{03Low}
\end{figure}

% DISCUTERE QUESTO PEZZO.....
As a last piece of evidence, we investigate the sensitivity of the previous results to the quench depth. 
 Figure \ref{03Low} shows the result for $\zeta=0.3$ following a quench from $T=6.0$ to $T=0.2$. 
 At this low $T$, we expect (based on extrapolation of the $T$-dependence of the equilibrium
 rotational and translational characteristic times) that also the $l=1$ correlator is unable to thermalize.
 Indeed, we find  that when $T_{bath}$ is low enough, the odd-rotations are frozen and hence coupled to the density, being described by the same $T_{eff}$.

\begin{figure}
\begin{center}
\includegraphics[width = 0.4\textwidth]{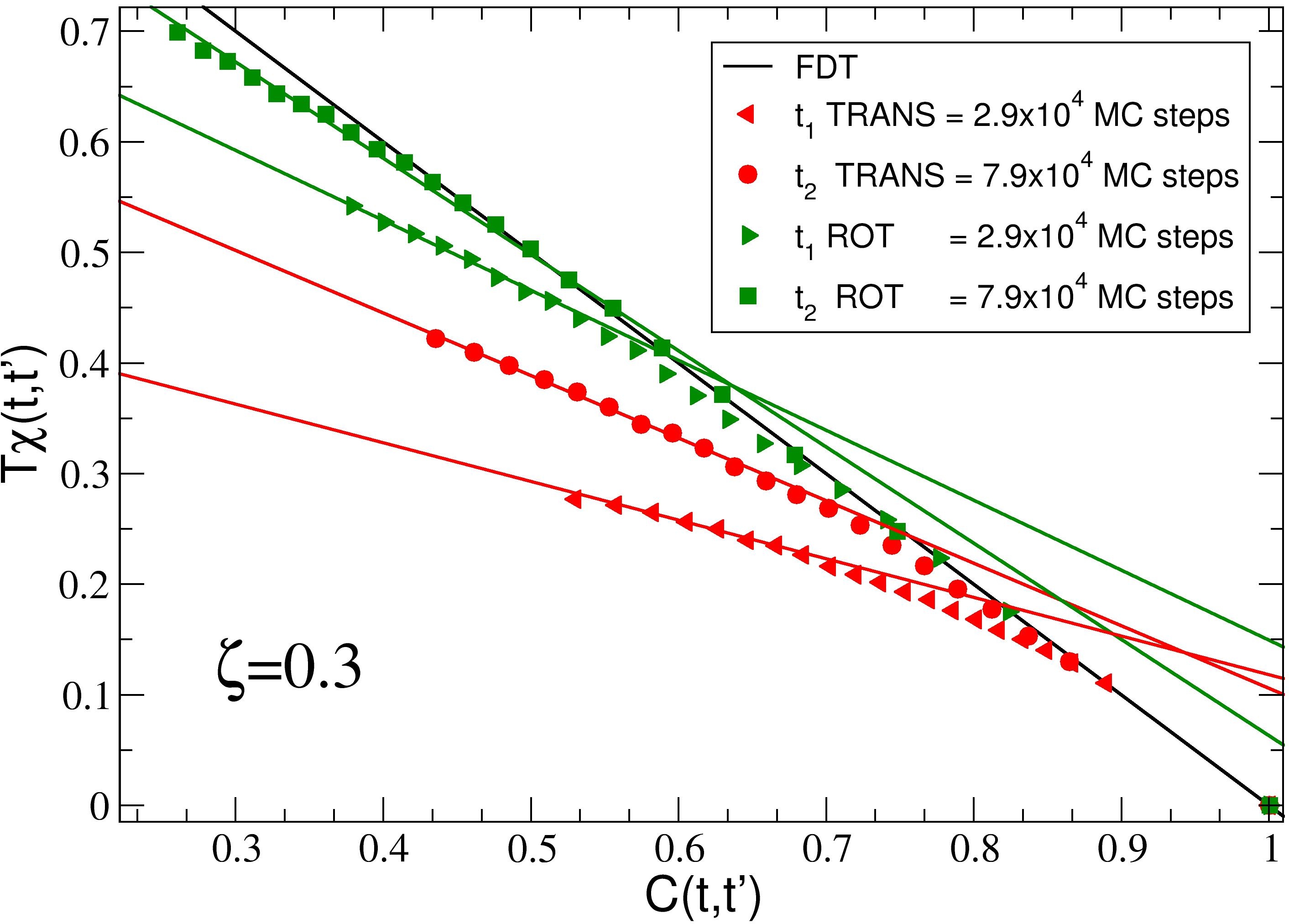}
\caption{self-density and $l=1$-orientation FD-plots for the system of dumbbells with $\zeta=0.3$ for a quench
from $T=6$ to $T=0.4$. Data plotted as (red) left and (green) right triangles are measured for a total time of $t=29400$ MC steps, while for (red) circles and (green) squares the total observation time is set to $t_2=79200$ MC steps.}
\label{compTw}
\end{center}
\end{figure}

Finally, to gain  insight into the off-equilibrium dynamics of molecular rotations, we study  the $t$ dependence of $T_{eff}$ by comparing  the correlators and the response functions for two different waiting time $t'$ for the quench $T=6.0\rightarrow T=0.4$.  
Figure \ref{compTw} shows the self-density and the $l=1$-orientation FD-plots for the observation times $t_1=29400$ and
$t_2=79200$ MC steps. 
We note that
for both translational and rotational degrees of freedom the  slope of the FD-plot  changes with $t'$, in agreement
with previous observations for atomic systems~\cite{Sciortino2001},  highlighting the presence of a slow aging process.
We also note that while the  $T_{eff}$ of the self-density  remains always very different from
the $T_{bath}$, the $l=1$-orientation appears to have almost completely thermalized at the largest $t'$ value.

\section*{Conclusions}
The idea of considering the $T_{eff}$ entering in the FD-ratio as a thermodynamic parameter that accounts for the aging
of the slow modes in structural glasses has been widely discussed in many different frameworks. One of the requests for
being a robust thermodynamic variable is its independence from the (slow) observable measured \cite{Cugliandolo2011}.
For this purpose we investigated the off-equilibrium dynamics of a van der Waals liquid composed of a binary mixture of dumbbell molecules, by studying the fluctuations and the responses of  observables associated both to translational
and rotational motions.  
Thus we have been able to carry on a more complete study with respect to the case of the atomic Lennard-Jones, where only the
wave-vector dependence of the density correlator~\cite{Berthier2004, Berthier2007MC} was investigated. 
In this article, we have reported the first computational measurements of the FDR for rotational degrees of freedom
in a molecular liquid. We note in passing that previous experimental studies of the FDT violation were built on the observation of 
rotational dynamics, based on depolarized light-scattering or dielectric spectroscopy \cite{Grigera1999,Buisson2005,Reiser2011,Oukris2010}.
The evaluation of $T_{eff}$ at different elongations, highlighted a rich and unexpected scenario. We focused on the off-equilibrium
dynamics of density and orientation fluctuations of the molecules and we observed that, for high elongations,
rotational and translational degrees of freedom are characterized by the same $T_{eff}$. Such situation is partially
lost at lower elongations.  For small elongations and shallow quenches,  odd rotational degrees of freedom are characterized by different 
FD-plots as compared to translation and the $T_{eff}$ of the rotational degrees of freedom 
approaches the bath temperature. For small elongations and  deep quenches, the effective temperatures
of the translational and rotational observables couple again. 

 These findings bear a resemblance with the $\zeta$ dependence of the equilibrium 
 rotational and translational behavior.  At small $\zeta$, molecules  undergo rotations of $180^\circ$ which allow $C_1(t)$ (but not $C_2(t)$) to relax fast \cite{Moreno2005},  decoupling translations and rotations. 
As a consequence, at equilibrium,  $C_1(t)$ does not show a separation of time-scales at
temperatures where $F_s(\textbf{k},t)$ and $C_2(t)$ are instead characterized by the typical two-step decay. 
Indeed,  in the limit of zero elongation, rotations are completely free and decoupled from translations.
Our results suggests that only at very deep temperatures,  odd-rotational dynamics  couples again
to translations. Indeed, only for deep quenches, the  $l=1$-orientation and the
self-density FD-plots display the same $T_{eff}$.  The  $l=2$-orientation instead is always coupled with
density.  We find that for shallow quenches, the  violation of the FDT for the $l=1$-orientation is a transient effect.
 Hence in a hypothetical experiment on molecular liquids with low elongation, the measurement of the
fluctuations and of the responses of observables coupled with degrees of freedom not sensitive to the hopping
processes  will  lead to a resulting $T_{eff}$ which  reflects the same aging behavior of  the  translational degrees of freedom. Differently, $l=1$ observables may or may not couple to   translational degrees of freedom.  The possibility thus exists that the effective temperature of the odd angular correlator might differ from the one of the translational
degrees of freedom. 
Finally, we want to stress that recent theoretical arguments \cite{Martens2009} provided evidence that, in off-equilibrium stationary states, the observable independence of $T_{eff}$ is related to the uniformity of the phase space distribution. It would be interesting to understand in future studies how our results connects with this appealing theoretical result.

\section*{acknowledgements}
N.G. acknowledges the support from MIUR-FIRB ANISOFT (RBFR125H0M). We thank A. Puglisi, A. Sarracino, and A. Crisanti for  discussions.

\appendix
\section{Comparison with Newtonian Dynamics}\label{appendix:MCvsMD}

In our simulation a Monte Carlo step is defined as $N$ attempts to  translate and rotate a molecule chosen randomly.
Translations are uniformly extracted  in the interval $[-\delta r_{MAX},\delta r_{MAX}]$ while rotations are performed
around a randomly chosen axis with an angle uniformly distributed within the interval
$[-\delta \alpha_{MAX},\delta \alpha_{MAX}]$. The attempt to move a molecule is rejected or accepted according to the
Metropolis rule \cite{Metropolis1953}. Hence the acceptance rate, and consequently the dynamics, depends on the balance
between the two parameters $\delta r_{MAX}$ and $\delta \alpha_{MAX}$. Since the acceptance rate has not to be neither
too small nor too high (i.e. should range between $30\%$ and $60\%$ of the total attempts) to provide a long-time dynamics
consistent with the Newtonian one, we set the acceptance rate in our simulation to $45\%$. Given this, there are a number of
combinations for $\delta r_{MAX}$ and $\delta \alpha_{MAX}$ that satisfy the condition on the 
acceptance rate. We then perform a number of simulations in the canonical ensemble where the two parameters are varied and we
observe the relaxation of density and angular correlation functions.
We report here the results for the intermediate state point $T=0.75$, $\phi=0.708$ when the elongation of the
dumbbell is set to $\zeta=0.5$, but the procedure to follow is the same at all the elongations.

\begin{figure*}[ht]
\begin{center}
\subfloat[]{\includegraphics[width = 0.4\textwidth]{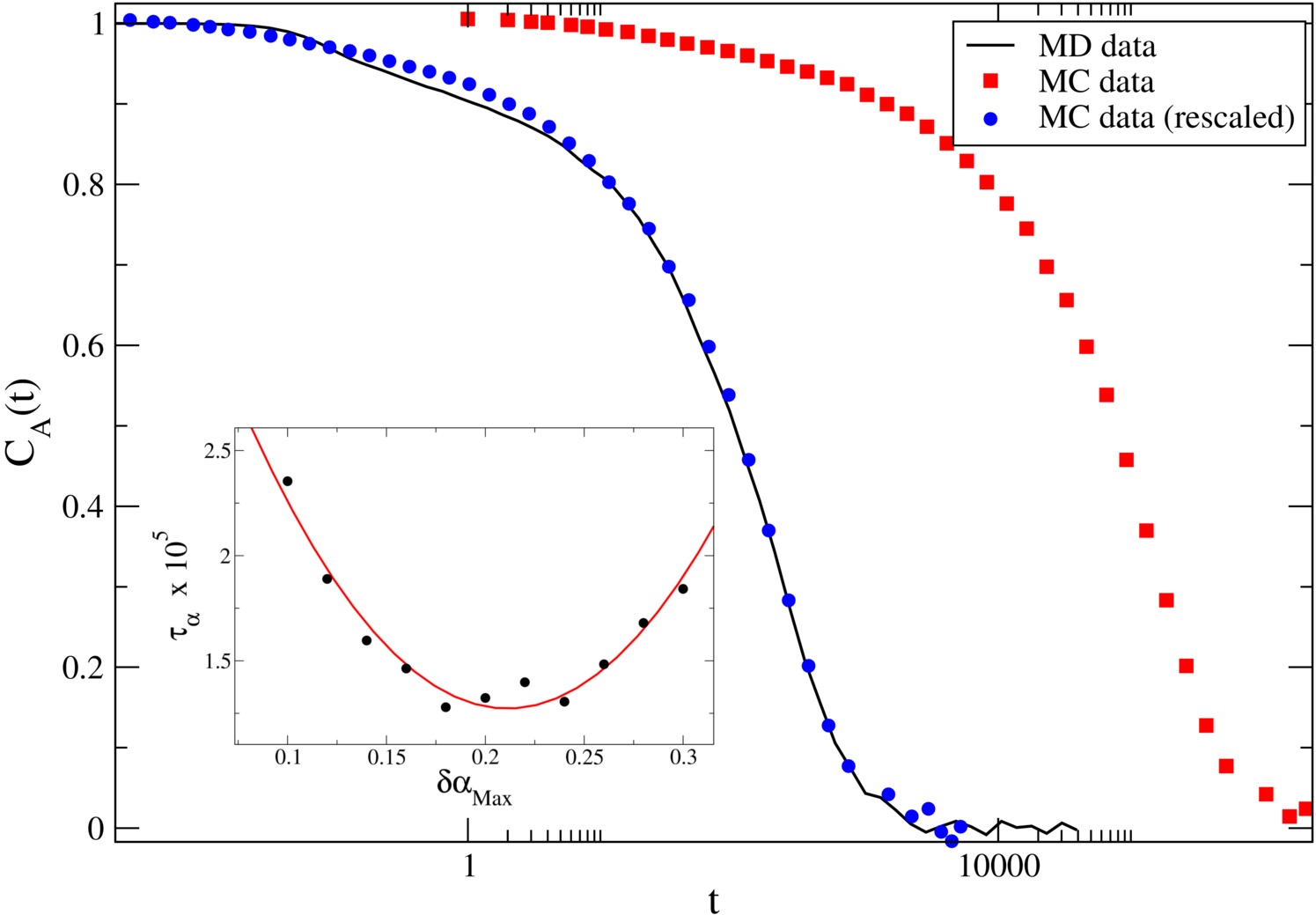}}\hspace{0.1cm}
\subfloat[]{\includegraphics[width = 0.4\textwidth]{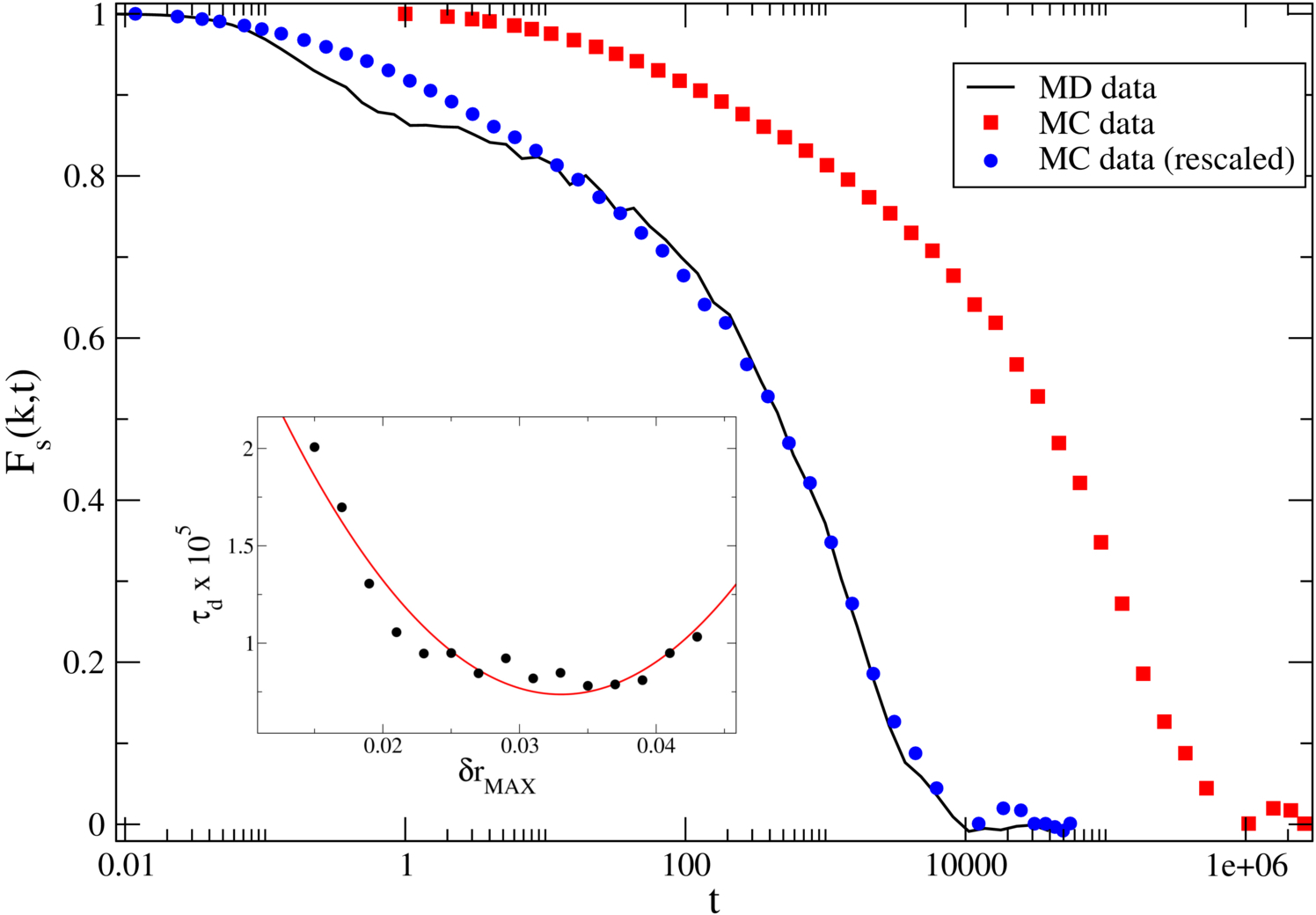}}
\caption{(a) Self-intermediate scattering function $F_s(\textbf{k},t)$ and (b) autocorrelation function $C_1(t)$ of the first Legendre polynomial for Dumbbell molecules with elongation $\zeta=0.5$ at the state point $T=0.75$,$\phi=0.708$. Symbols are results from MC dynamics and solid lines from Newtonian dynamics (data taken from Ref.\cite{Chong2005}).
Insets: evolution of the relaxation times (a)for $F_s(k,t)$ as a function of $\delta r_{MAX}$ (b) for $C_{1}(t)$ as a function of $\delta \alpha_{MAX}$. The minimum values of $\delta r_{MAX}$ and $\delta \alpha_{MAX}$ used in MC simulations provide a relaxation behavior of functions in (a) and (b), compatible with the results obtained in Molecular dynamic simulations.}
\label{MDMCango}
\end{center}
\end{figure*}

We extract from simulations the relaxation times of the self-intermediate scattering function $\tau_d$ and of the
angular correlator $ \tau_{\alpha}$, defined as the value at which the correlation
functions decay at $1/e$. The relaxation times  (in unit of MC step)  $\tau_{\alpha}$ and  $\tau_d$ 
as a function of  $\delta r_{MAX}$ and $\delta \alpha_{MAX}$ 
are shown in the insets of fig.\ref{MDMCango} (a) and (b) respectively. Due to their non-monotonic behaviour we are able to extrapolate through a parabolic fit the value at which $\tau_d$ and $\tau_{\alpha}$ are minimized, finding $\delta r_{MAX}=0.025$ and $\delta \alpha_{MAX}=0.20$. Figure \ref{MDMCango} (a) and (b) show that for such values, the long-time decay of the angular correlation function and of the self-intermediate scattering function  can be superimpose on top of the corresponding correlators obtained from Newtonian dynamics, after having scaled the $x$-axis of the MC curves by an arbitrary factor.  

\newpage
\bibliography{library}

\begin{thebibliography}{10}

\bibitem{GOETZE2009}
W. Goetze, {\em {Complex dynamics of glass-forming liquids: a mode coupling
  theory}}, 1st ed. (Oxford University Press, Oxford, 2009).

\bibitem{BINDER2005}
K. Binder and W. Kob, {\em {Glassy materials and disordered solids: an
  introduction to thei statistical mechanics}}, 1st ed. (World Scientific,
  Singapore, 2005).

\bibitem{WOLYNES2012}
P.~G. Wolynes and V. Lubchenko, {\em {Structural glasses and supercooled
  liquids}}, 1st ed. (Wiley, Hoboken, New Jersey, 2012).

\bibitem{Leuzzi2009}
L. Leuzzi, J. Non-Cryst. Solids {\bf 355},  686  (2009).

\bibitem{Berthier2011}
L. Berthier and G. Biroli, Rev. Mod. Phys. {\bf 83},  587  (2011).

\bibitem{Biroli2013}
G. Biroli and J.~P. Garrahan, J. Chem. Phys. {\bf 138},  12A301
  (2013).

\bibitem{Kurchan2005}
J. Kurchan, Nature {\bf 433},  222  (2005).

\bibitem{Ediger1996}
M.~D. Ediger, C.~A. Angell, and S.~R. Nagel, J. Chem. Phys. {\bf
  100},  13200  (1996).

\bibitem{Kovacs1963}
A.~J. Kovacs, Fortschr. Hochpolym.-Forsch. {\bf 3},  394  (1963).

\bibitem{Angell2000}
C.~a. Angell {\it et~al.}, J. of Appl. Phys. {\bf 88},  3113  (2000).

\bibitem{Mossa2004}
S. Mossa and F. Sciortino, Phys. Rev. Lett. {\bf 92},  045504  (2004).

\bibitem{Saika2004}
I. Saika-Voivod and F. Sciortino, Phys. Rev. E {\bf 70},  041202  (2004).

\bibitem{Cugliandolo1997}
L.~F. Cugliandolo, J. Kurchan, and L. Peliti, Phys. Rev. E {\bf 55},  3898
   (1997).

\bibitem{Cugliandolo2000}
L. Cugliandolo and J. Kurchan, J. Phys. Soc. Japan, Suppl.A {\bf 69},  247
  (2000).

\bibitem{Crisanti2003}
A. Crisanti and F. Ritort, J. Phys. A {\bf 36},  R181  (2003).

\bibitem{Cugliandolo2011}
L. Cugliandolo, J. Phys. A {\bf 44},  4830001  (2011).

\bibitem{Parisi1997}
G. Parisi, Phys. Rev. Lett. {\bf 79},  19 (1997).

\bibitem{Barrat1999a}
J.-l. Barrat and W. Kob, Europhys. Lett. {\bf 46},  637  (1999).

\bibitem{Kob2000}
W. Kob and J. Barrat, Eur. Phys. J. B {\bf 13},  319  (2000).

\bibitem{Leonardo2000}
R.~D. Leonardo, L. Angelani, G. Parisi, and G. Ruocco, Phys. Rev. Lett.
  {\bf 84},  6054  (2000).

\bibitem{Sciortino2001}
F. Sciortino and P. Tartaglia, Phys. Rev. Lett. {\bf 86},  107  (2001).

\bibitem{Berthier2007MC}
L. Berthier and W. Kob, J. Phys.: Condens. Matt. {\bf 19},  205130
  (2007).

\bibitem{Gnan2010}
N. Gnan, C. Maggi, T. Schroeder, and J. Dyre, Phys. Rev. Lett. {\bf
  104},  125902  (2010).

\bibitem{Gnan2013}
N. Gnan, C. Maggi, G. Parisi, and F. Sciortino, Phys. Rev. Lett. {\bf 110},
  035701  (2013).

\bibitem{Sciortino2005}
F. Sciortino, J. Stat. Mech. Theor. Exp. {\bf
  2005},  P05015  (2005).

\bibitem{Pedersen2008}
U.~R. Pedersen, T. Christensen, T.~B. Schroeder, and J. Dyre, Phys. Rev. E
  {\bf 77},  011201  (2008).

\bibitem{Bailey2008}
N.~P. Bayley {\it et~al.}, J. Chem. Phys. {\bf 129},  184507
  (2008).

\bibitem{Berthier2004Shear}
L. Berthier and J.-L. Barrat, J. Chem. Phys. {\bf 116},  6228
  (2004).

\bibitem{Berthier2007a}
L. Berthier, Phys. Rev. Lett. {\bf 98},   220601  (2007).

\bibitem{Grigera1999}
T.~S. Grigera and N.~E. Israeloff, Phys. Rev. Lett. {\bf 83},  5038
  (1999).

\bibitem{Buisson2005}
L. Buisson and S. Ciliberto, Physica D {\bf 204},  1
  (2005).

\bibitem{Reiser2011}
J. Schindele, A. Reiser, and C. Enss, Phys. Rev. Lett. {\bf 107},
  095701  (2011).

\bibitem{Oukris2010}
H. Oukris and N.~E. Israeloff, Nature Physics {\bf 6},  135  (2010).

\bibitem{Chong2005}
S.-H. Chong, A.~J. Moreno, F. Sciortino, and W. Kob, Phys. Rev. Lett.
  {\bf 4},  94  (2005).

\bibitem{kobAndersen1995}
H.~C. Andersen and W. Kob, Phys. Rev. E {\bf 51},  5  (1995).

\bibitem{kobAndersen19952}
H.~C. Andersen and W. Kob, Phys. Rev. E {\bf 52},  4  (1995).

\bibitem{Moreno2005}
A.~J. Moreno, S.-H. Chong, W. Kob, and F. Sciortino, The Journal of chemical
  physics {\bf 123},  204505  (2005).

\bibitem{Sanz2010}
E. Sanz and D. Marenduzzo, J. Chem. Phys. {\bf 132},  194102
  (2010).

\bibitem{Romano2011}
F. Romano, C. De~Michele, D. Marenduzzo, and E. Sanz, Journal of Chemical
  Physics {\bf 135},  124106  (2011).

\bibitem{Gleim1998}
T. Gleim {\it et~al.}, Phys. Rev. Lett. {\bf 81},  4404  (1998).

\bibitem{Barrat1999b}
J.-l. Barrat and W. Kob, J. Phys.  Condens. Matt. {\bf 11},  A247
  (1999).

\bibitem{Chong2002}
S.-H. Chong and W. G\"{o}tze, Phys. Rev. E {\bf 65},  041503  (2002).

\bibitem{Berthier2004}
L. Berthier, Phys. Rev. E {\bf 69},  020201  (2004).

\bibitem{Martens2009}
K. Martens, E. Bertin, and M. Droz, Phys. Rev. Lett. {\bf 3 },  260602
  (2009).

\bibitem{Metropolis1953}
N. Metropolis {\it et~al.}, J. Chem. Phys. {\bf 21},  1087
  (1953).

\end{thebibliography}
\end{document}